\documentclass[a4paper,11pt]{article}
\usepackage{jheppub}

\usepackage{amsmath,epsfig}
\usepackage{lscape}
\usepackage{multirow}

\title{Particle yields in heavy ion collisions and the influence of strong magnetic fields}

\author{M.G. de Paoli}
\author{and D.P. Menezes}
\affiliation{Depto de F\'{\i}sica - CFM - Universidade Federal de Santa Catarina
\\Florian\'opolis - SC - CP. 476 - CEP 88.040 - 900 - Brazil}

\emailAdd{marcelodepaoli@gmail.com}
\emailAdd{debora.p.m@ufsc.br}

\abstract{It is expected that the magnetic field in the surface of magnetars do not exceed 
$10^{15}$ G. However, in heavy ion collisions, this value is expected
to be much higher.
We investigate the effects of a magnetic fields varying from $10^{18}$, 
to $10^{20}$ G in strange matter (composed of $u$, $d$ and $s$ 
quarks).
We model matter as a free gas of baryons and mesons under the influence of an 
external magnetic field. We study the effects of such strong fields 
through a $\chi^2$ fit to some data sets of the STAR experiment.
For this purpose we solve the Dirac, Rarita-Schwinger, Klein-Gordon and Proca
equations subject to magnetic fields in order to obtain the energy 
expressions and the degeneracy for spin $1/2$, spin $3/2$, spin $0$ and 
spin $1$ particles, respectively.
Our results show that a field of the order of $10^{19}$ G produces
an improved fitting
to the experimental data as compared to the calculations without magnetic 
field.}

\begin{document}
\maketitle
\flushbottom

\section{Introduction}

According to Quantum Chromodynamics, the quark-gluon plasma (QGP) phase 
refers to matter where quarks and gluons are believed to be deconfined and it 
probably takes place at temperatures of the order of 150 to 170 MeV. 
In large  colliders around the world 
(RHIC/BNL, ALICE/CERN, GSI, etc), physicists are trying to
find a QGP signature looking at  
non-central heavy ion collisions.

Possible experiments towards this search are Au-Au collisions at RHIC/BNL
and Pb-Pb collisions at SPS/CERN, where the
hadron abundances and particle ratios are used in order to determine the 
temperature and baryonic chemical potential of the possibly present 
hadronic matter-QGP phase transition.

In previous papers a statistical model under chemical equilibration was 
used to calculate particle yields \cite{munzinger,munzinger2} and in these
works the densities of particles were obtained from free Fermi and Boson gas 
approximations, where the interaction among the baryons and mesons were 
neglected. More recently, relativistic nuclear models have been tested in the 
high temperature regime produced in these heavy ion collisions. 
In \cite{nosso1,nosso2} different versions of Walecka-type relativistic models
\cite{sw} were used to calculate the Au-Au collision particle yields at RHIC/BNL
and in \cite{nosso_qmc} the quark-meson-coupling model \cite{qmc,qmc2,qmc3} was used
to calculate this reaction results and also Pb-Pb collision particle rations 
at SPS/CERN.
In all cases 18 baryons, pions, kaons, $\rho$'s and $K^*$s were incorporated 
in the calculations and a fit based on the minimum value of the
quadratic deviation was implemented in
order to obtain the temperature and chemical potential for each model,
according to a prescription given in \cite{munzinger}.
For Au-Au collision (RHIC) these numbers lie in the range 
$132 < T < 169$ MeV and $30.5 < \mu_B < 62.8$ MeV  and for Pb-Pb
collision (SPS),
$99 < T < 156.1$ MeV and $167.5 < \mu_B < 411$ MeV. 

On the other hand, the magnetic fields involved in heavy-ion collisions
\cite{kharzeev,kharzeev2,kharzeev3} can reach 
intensities even higher than the ones considered in magnetars 
\cite{magnetars,magnetars2}. As suggested in \cite{kharzeev,kharzeev2,kharzeev3} and 
\cite{eduardo,eduardo2,eduardo3} it is interesting to investigate fields of the order 
of $eB=5 -30 m_\pi^2$ (corresponding to $1.7 \times 10^{19} - 10^{20}$ Gauss)
and temperatures varying from $T=120-200$ MeV related to heavy ion collisions.
In fact, the densities related to the chemical potentials obtained within
the relativistic models framework, in all cases, are very low (of the
order of $10^{-3}$ fm$^{-3}$). At these densities the nuclear interactions are
indeed very small and this fact made us reconsider the possibility of free
Fermi and Boson gases, but now under the influence of strong magnetic 
fields.

 In a recent paper \cite{tuchin}, the author studies the
 synchrotron radiation of gluons by fast quarks in strong magnetic
 fields produced in heavy ion collisions and shows that a strong
 polarization of quarks and leptons with respect to the direction of
 the magnetic field is expected. The polarization of quarks seems 
 to be washed out during the fragmentation but this is not
 the case of the leptons. The  observation of lepton polarization asymmetry 
 could be a proof of the existence of the magnetic field,
 which may last for $1-2$ fm/c. This slowly varying magnetic field could 
 leave its signature in the particle yields.

 The purpose of the analysis we present in this paper is to check
  if the inclusion of strong magnetic fields can improve the fitting
  of experimental results. We start from the simplest possible
  calculation, assuming that the magnetic field is homogeneous,
  constant and time-independent. We are aware that it is not the case,
as shown in \cite{skokov,deng}, where the shape of
the magnetic field presents a special non-trivial pattern. 
Moreover, from the calculations performed in these references, one
can see that after averaging over many events one is left with
just of the components of the magnetic field. Nevertheless, the
event-by-event fluctuation of the position of charged particles can
induce another component of the magnetic field (perpendicular to the
remaining one in the average calculation) and also an
electric field, which is quite strong at low impact
parameters. While the magnetic field remains quite high in peripheral collisions,
the opposite happens with the electric field.
To make our first analysis as simple as possible, we shall restrict
ourselves to data at centralities of the order of 80$\%$, i.e., high values of the impact
parameter $b \simeq 11-13$ fm, where we are more comfortable to
disregard the electric field effects.

In the present paper we briefly revisit the formalism necessary for the 
calculation of particle densities subject to magnetic fields and the expressions
used to implement a $\chi^2$ fit to the experimental results.

\section{Formalism}

We model matter as a free gas of baryons and mesons under the influence of a constant magnetic field.
We consider only normal and strange matter, i.e., the baryons and mesons constituted by $u$, $d$ and $s$ quarks:
the baryon octet (spin 1/2 baryons), the baryon decuplet (spin 3/2 baryons), the pseudoscalar meson nonet (spin 0 mesons) and
the vector meson nonet (spin 1 mesons),
which leaves us with a total of $54$ particles ($18$ baryons, $18$ antibaryons and $18$ mesons).

We utilize natural units ($\hbar=c=1$) and define $\epsilon_0=\mu_0=1$. From 
the relation
$\alpha=\frac{e^2}{4\pi \epsilon_0 \hbar c}$ we obtain that the electron charge is $e=\sqrt{4\pi\alpha}$,
where $\alpha=\frac{1}{137}$ is the fine structure constant.
The natural units with the electron charge in that form is known as 
Heaviside-Lorentz units \cite{jackson}.

In this work, the magnetic field is introduced trough minimal coupling, so 
the derivatives become
\begin{equation}
 \partial_\mu \rightarrow D_\mu=\partial_\mu +i q A_\mu.
\end{equation}

We write the charge as $q=\epsilon_q|q|$, where $\epsilon_q=+(-)$ corresponds
to a particle with positive (negative) charge, and assume the gauge
\begin{equation}
 A_\mu=\delta_{\mu 2} x_1 B \quad \rightarrow \quad A_0=0 \quad \text{and} \quad \vec{A}=(0,x_1 B,0),
\end{equation}
so,
\begin{equation}
 \vec{\nabla} \cdot \vec{A}=0 \quad \text{and} \quad \vec{\nabla} \times \vec{A}=B\hat{e}_3,
\end{equation}
and the derivatives
\begin{equation}
  D_\mu=\partial_\mu -i\epsilon_q|q|Bx_1\delta_{\mu 2}.
\end{equation}

We search for solutions of the fields $\psi$ in the form
\begin{equation}
  \psi^{(\epsilon)}_\alpha=
\left\{
\begin{matrix}
C^{(\epsilon)}_\alpha e^{-i\epsilon Et+i \epsilon \vec{p}\cdot \vec{x}}           & (q=0)\\
f^{(\epsilon)}_\alpha(x_1)e^{-i\epsilon Et+i \epsilon p_2 x_2+i\epsilon p_3 x_3} & (q\neq0)
\end{matrix}
\right.,
\end{equation}
where $\psi_\alpha$ are the components of the field $\psi$ and $\epsilon=+(-)$ corresponds to the states of positive (negative) energy.

For the spin 1/2 baryons (Dirac field) $\psi$ has $4$ components,
for the spin 3/2 baryons (Rarita-Schwinger field) $\psi_\mu$ has $16$ components, 
for the spin 0 mesons (Klein-Gordon field) $\psi$ has just one component,
and for the spin 1 mesons (Proca field) $\psi_\mu$ has $4$ components.

Due to the use of statistical methods to deal with the system under 
consideration, we do not need the complete expression for $\psi$, but
just the form of the energy $E$ for each one of the fields and 
the degeneracy of the energy levels $\gamma$.

\subsection{Spin 1/2 Baryons}

The baryons with spin 1/2 are described by the Dirac Lagrangian density 
\cite{melrose}
\begin{equation}
{\cal L}^D=\bar{\psi}(i\gamma^\mu D_\mu-m)\psi,
\end{equation}
which (after we apply the Euler-Lagrange equation) lead us to the equation of motion
\begin{equation}
 (i\gamma^\mu D_\mu - m)\psi=0.
\end{equation}
where $\gamma^\mu$ are the Dirac matrices.

The solution of the equation of motion gives
\begin{equation}
E=\left\{
\begin{matrix}
 &\sqrt{\vec{p}^2+m^2}    & (q=0) \\
 &\sqrt{p_3^2+m^2+2\nu|q|B} & (q\neq0)
\end{matrix}
\right.,
\end{equation}
where $\nu$ runs over the possible Landau Levels and the degeneracy for the 
energy states are given by:
\begin{equation}
\gamma=\left\{
 \begin{matrix}
  &2             &(q=0) \\
  &2-\delta_{\nu0} &(q\neq0)
 \end{matrix}
\right..
\end{equation}

\subsection{Spin 3/2 Baryons}

The baryons with spin 3/2 are described by the Rarita-Schwinger Lagrangian
density \cite{rs,weinberg}
\begin{equation}
{\cal L}^{RS}=-\frac{1}{2}\bar{\psi}_\mu(\epsilon^{\mu\nu\rho\sigma} \gamma_5 \gamma_\nu D_\rho+im\sigma^{\mu\sigma})\psi_\sigma,
\end{equation}
where $\gamma^5=i\gamma^0\gamma^1\gamma^2\gamma^3$ and $\sigma^{\mu\nu}=\frac{i}{2}[\gamma^\mu,\gamma^\nu]$.

The equation of motion reads
\begin{equation}
 (i\gamma^\mu D_\mu - m)\psi_\nu=0
\quad \text{with} \quad
 \gamma^\mu\psi_\mu=0 \quad \text{and} \quad D^\mu\psi_\mu=0.
\end{equation}

The solution of the Rarita-Schwinger equation is not trivial and poses
non-causality problems. To obtain the degeneracy of the energy
states, we follow the prescription used in \cite{melrose}, which
is given in detail for the Rarita-Schwinger equation in \cite{ours}.
Observing the equation of motion one can see that each component of $\psi_\mu$ obeys a Dirac type equation,
so the energy must have the form
\begin{equation}
E=\left\{
\begin{matrix}
 &\sqrt{\vec{p}^2+m^2}    & (q=0) \\
 &\sqrt{p_3^2+m^2+2\nu|q|B} & (q\neq0)
\end{matrix}
\right..
\end{equation}

Besides that, $\psi_\mu$ has $4$ components, but, two equations are 
constrainted, which means that only $2$ components of $\psi_\mu$ are really 
independent.
So, $\psi_\mu$ have $2$ polarizations, but, (because of the Dirac equation solution) each polarization are double degenerate.
In the presence of a magnetic field
there is another constraint for the $\nu=0$ and $\nu=1$ energy levels,
which leads to the following degeneracy for the energy states
\begin{equation}
\gamma=\left\{
 \begin{matrix}
  &4             &(q=0) \\
  &4-2\delta_{\nu 0} -\delta_{\nu 1}  &(q\neq0)
 \end{matrix}
\right..
\end{equation}

\subsection{Spin 0 Mesons}

The mesons with spin 0 are described by the Klein-Gordon Lagrangian density
\cite{greiner}
\begin{equation}
{\cal L}^{KG}=D^\mu \psi^\ast D_\mu \psi - m^2\psi^\ast\psi,
\end{equation}
whose equation of motion is given by
\begin{equation}
(D^\mu D_\mu+m^2)\psi=0,
\end{equation}
with the energy satisfying the relation:
\begin{equation}
E=\left\{
\begin{matrix}
 &\sqrt{\vec{p}^2+m^2}    & (q=0) \\
 &\sqrt{p_3^2+m^2+(2\nu+1)|q|B} & (q\neq0)
\end{matrix}
\right..
\end{equation}

\subsection{Spin 1 Mesons}

The mesons with spin 1 are described by the Proca Lagrangian density 
\cite{russo}
\begin{equation}
{\cal L}^{P}=\frac{1}{2}(D^\mu\psi^{\nu \ast}-D^\nu\psi^{\mu \ast})(D_\mu\psi_\nu-D_\nu\psi_\mu)-m^2\psi^{\nu \ast}\psi_\nu.
\end{equation}

The equation of motion is
\begin{equation}
(D^\mu D_\mu+m^2)\psi_\nu=0 \quad \text{with} \quad D_\mu\psi^\mu=0.
\end{equation}

Each component of $\psi_\mu$ obey a Klein-Gordon type equation, so that 
the energy states are
\begin{equation}
E=\left\{
\begin{matrix}
 &\sqrt{\vec{p}^2+m^2}    & (q=0) \\
 &\sqrt{p_3^2+m^2+(2\nu+1)|q|B} & (q\neq0)
\end{matrix}
\right.,
\end{equation}
$\psi_\mu$ has $4$ components, but, one of the equations is a compressed constraint equation,
which means that only $3$ components of $\psi_\mu$ are independent.
So, each energy state have $3$ polarizations in the case with zero charge 
(or without magnetic field).
If the charge is different from zero (and we have the presence of an 
external magnetic field) there is an additional constraint for the 
$\nu=0$ energy level,
which leads to the following degeneracy for the energy states
\begin{equation}
\gamma=\left\{
 \begin{matrix}
  &3             &(q=0) \\
  &3-\delta_{\nu0} &(q\neq0)
 \end{matrix}
\right..
\end{equation}

\subsection{Thermodynamics}

Using the Grand Canonical formalism we obtain that the particle densities 
for the baryons are
\begin{equation}
 \rho_b=\left\{
\begin{aligned}
 \gamma_b \frac{1}{2\pi^2}\int^\infty_0f(E_b-\mu_b)p^2dp \quad (q=0) \\
  \sum_{\nu=0}^\infty \gamma_b \frac{|q_b|B}{2\pi^2}\int^\infty_0f(E_b-\mu_b)dp \quad (q\neq0)
\end{aligned}
\right.,
\end{equation}
for the antibaryons are
\begin{equation}
 \rho_{ab}=\left\{
\begin{aligned}
 \gamma_b \frac{1}{2\pi^2}\int^\infty_0f(E_b+\mu_b)p^2dp \quad (q=0) \\
  \sum_{\nu=0}^\infty \gamma_b \frac{|q_b|B}{2\pi^2}\int^\infty_0f(E_b+\mu_b)dp \quad (q\neq0)
\end{aligned}
\right.,
\end{equation}
and for the mesons are
\begin{equation}
 \rho_m=\left\{
\begin{aligned}
 \gamma_m \frac{1}{2\pi^2}\int^\infty_0b(E_m-\mu_m)p^2dp \quad (q=0) \\
  \sum_{\nu=0}^\infty \gamma_m \frac{|q_m|B}{2\pi^2}\int^\infty_0b(E_m-\mu_m)dp \quad (q\neq0)
\end{aligned}
\right.,
\end{equation}
with $f(x)=(e^{x/T}+1)^{-1}$ and $b(x)=(e^{x/T}-1)^{-1}$.

The total baryonic particle density is
\begin{equation}
 \rho_B=\sum_b(\rho_b-\rho_{ab}),
\end{equation}
and the total mesonic density is
\begin{equation}
 \rho_M=\sum_m\rho_m.
\end{equation}

The energy density is given by the sum of the energy densities 
of each particle, so
\begin{equation}
 \epsilon=\sum_b(\epsilon_b+\epsilon_{ab})+\sum_m\epsilon_m,
\end{equation}
with
\begin{equation}
 \epsilon_b=\left\{
\begin{aligned}
 \gamma_b \frac{1}{2\pi^2}\int^\infty_0 E_b f(E_b-\mu_b)p^2dp \quad (q=0) \\
  \sum_{\nu=0}^\infty \gamma_b \frac{|q_b|B}{2\pi^2}\int^\infty_0 E_b f(E_b-\mu_b)dp \quad (q\neq0)
\end{aligned}
\right.,
\end{equation}
\begin{equation}
 \epsilon_{ab}=\left\{
\begin{aligned}
 \gamma_b \frac{1}{2\pi^2}\int^\infty_0 E_b f(E_b+\mu_b)p^2dp \quad (q=0) \\
  \sum_{\nu=0}^\infty \gamma_b \frac{|q_b|B}{2\pi^2}\int^\infty_0 E_b f(E_b+\mu_b)dp \quad (q\neq0)
\end{aligned}
\right.,
\end{equation}
\begin{equation}
 \epsilon_m=\left\{
\begin{aligned}
 \gamma_m \frac{1}{2\pi^2}\int^\infty_0 E_m b(E_m-\mu_m)p^2dp \quad (q=0) \\
  \sum_{\nu=0}^\infty \gamma_m \frac{|q_m|B}{2\pi^2}\int^\infty_0 E_m b(E_m-\mu_m)dp \quad (q\neq0)
\end{aligned}
\right.,
\end{equation}
in the same way the pressure is given by
\begin{equation}
 P=\sum_b (P_b+P_{ab})+\sum_m P_m,
\end{equation}
with
 \begin{equation}
 P_b=\left\{
\begin{aligned}
 \gamma_b \frac{1}{6\pi^2}\int^\infty_0 \frac{1}{E_b}f(E_b-\mu_b)p^4dp \quad (q=0) \\
  \sum_{\nu=0}^\infty \gamma_b \frac{|q_b|B}{2\pi^2}\int^\infty_0 \frac{1}{E_b} f(E_b-\mu_b)p^2dp \quad (q\neq0)
\end{aligned}
\right.,
\end{equation}
 \begin{equation}
 P_{ab}=\left\{
\begin{aligned}
 \gamma_b \frac{1}{6\pi^2}\int^\infty_0 \frac{1}{E_b}f(E_b+\mu_b)p^4dp \quad (q=0) \\
  \sum_{\nu=0}^\infty \gamma_b \frac{|q_b|B}{2\pi^2}\int^\infty_0 \frac{1}{E_b} f(E_b+\mu_b)p^2dp \quad (q\neq0)
\end{aligned}
\right.,
\end{equation}
 \begin{equation}
 P_m=\left\{
\begin{aligned}
 \gamma_m \frac{1}{6\pi^2}\int^\infty_0 \frac{1}{E_m}b(E_m-\mu_m)p^4dp \quad (q=0) \\
  \sum_{\nu=0}^\infty \gamma_m \frac{|q_m|B}{2\pi^2}\int^\infty_0 \frac{1}{E_m} b(E_m-\mu_m)p^2dp \quad (q\neq0)
\end{aligned}
\right.,
\end{equation}
the entropy density can be found through
\begin{equation}
 s=\epsilon+P-\sum_b \mu_b (\rho_b-\rho_{ab})-\sum_m\mu_m\rho_m.
\end{equation}

\subsection{Chemical Potential}

The hadron chemical potential is
\begin{equation}
 \mu_h=B_h\;\mu_B+I_{3h}\;\mu_{I_3}+S_h\;\mu_S,
\end{equation}
where $B_h$, $I_{3h}$ and $S_h$, are respectively the baryonic number, the third isospin component and the strangeness of the particle $h$.
The baryonic chemical potential $\mu_B$ is a free parameter of the system (the other is the temperature $T$).
The chemical potential of isospin $\mu_{I_3}$ and strangeness $\mu_S$ are determined trough their respectively conservation laws.

We impose the local conservation of the baryonic number, isospin and strangeness. This imposition leads to the following equations
\begin{equation}
\sum_h B_h\;\rho_h=\frac{N_B}{V},
\quad
\sum_h I_{3h}\;\rho_h=\frac{I_3}{V},
\quad
\sum_h S_{h}\;\rho_h=\frac{S}{V},
\end{equation}
where $N_B$ is the total baryonic number, $I_3$ is the total isospin,
$S$ is the total strangeness of the system and $V$ are
the volume occupied by the system.
The charge conservation is automatically achieved trough the other three conservation laws.

The baryonic number of an Au atom is $N_B=(N+Z)=79+118=197$, the isospin is $I_3=(Z-N)/2=19.5$
and for the deuteron ($d$) we have that $N_B=1+1=2$ and $I_3=0$.
Hence, assuming that the total strangeness of the system is zero, we write the following table for the conserved quantities:

\begin{itemize}
 \item{Au$+$Au Collision, $N_B=394$, $I_3=-39$, $S=0$.} 
 \item{$d+$Au Collision, $N_B=199$, $I_3=-19.5$, $S=0$.}
\end{itemize}

At this point it is important to emphasize some of the drawbacks
of our simple calculation. As shown in \cite{kharzeev_ref}, the 
magnetic field should depend on the charges of the colliding nuclei 
and the number of participants should vary for different centralities.
These constraints were not taken into account directly in our
calculations. All the information we use as input come from the
experimental particle yields and the magnetic field is modified until
the best fitting is encountered. The number of
different participants is reflected only in the resulting radii.

\section{Results and Discussions}

We have  implemented a $\chi^2$ fit in
order to obtain the temperature and chemical potential.
The particle properties (spin, mass, baryonic number, isospin 
and strangeness) were taken from the \textit{Particle Data Group} \cite{rpp-2010}.

In tables 1, 2, 3 and 4  we show our results
corresponding to the temperature and chemical potential that
give the minimum value for the quadratic deviation $\chi^2$:
\begin{equation}
\chi^2 = \sum_i \frac{({\cal R}_i^{exp} -{\cal R}_i^{theo})^2}
{\sigma_i^2},
\end{equation}
where ${\cal R}_i^{exp}$ and ${\cal R}_i^{theo}$ are the $i^{th}$ particle
ratio given experimentally and theoretically, and $\sigma_i$
represents the errors in the experimental data points.

To make clear the improvement in the data fitting by the addition of the magnetic field,
we calculate the relative percent deviation $(\Delta_\%)$ with respect
to the experimental values for $B=0$ and the best $B\neq0$ (the bold columns in the tables)
trough the equation
\begin{equation}
 \Delta_\%=\Bigg|\frac{{\cal R}^{theo}-{\cal R}^{exp}}{{\cal R}^{exp}}\Bigg|\cdot100\%,
\end{equation}
and show these values in parenthesis in all the tables.

For the simulations our code deals with $5$ unknowns 
($\mu_B$, $\mu_{I3}$, $\mu_S$, $T$, $V$)
and $3$ constrained equations. We run over the values of $\mu_B$ and 
$T$ (the free parameters) in order to find the smallest $\chi^2$.
Our results are given next.

In tables 1, 2, 3 and 4, $B$ is the magnetic field, $T$ is the temperature, $\mu_B$ is the baryonic chemical potential,
$\chi^2$ is quadratic deviation, $\mu_{I3}$ is the isospin chemical potential, $\mu_S$ is the strangeness chemical potential,
$R$ is the radius of the "fire-ball",
$\rho_B=\sum_b(\rho_b-\rho_{ab})$ is the usual baryonic density,
$\rho_\Delta=\rho_{\Delta^{++}}-\rho_{\bar\Delta^{++}}+\rho_{\Delta^+}-\rho_{\bar\Delta^+}
+\rho_{\Delta^0}-\rho_{\bar\Delta^0}+\rho_{\Delta^-}-\rho_{\bar\Delta^-}$ is delta baryon density,
$\rho_M=\sum_m\rho_m$ is the meson density,
$\rho_\pi=\rho_{\pi^0}+\rho_{\pi^+}+\rho_{\pi^-}$ is the pion density,
$\epsilon$ is the energy density, $P$ is the pressure, $s$ is the entropy density and
$ndf$ is the number of degrees of freedom.
For $B=0$, $ndf=5$ ($7$ experimental values minus $2$ free parameters, $T$ and $\mu$),
for $B\neq0$, $ndf=4$ ($7$ experimental values minus $3$ free parameters, $T$, $\mu$ and $B$).
$\pi^-/\pi^+$, $K^-/K^+$, $\bar{p}/p$, $K^-/\pi^-$, $K^+/\pi^+$ and $p/\pi^+$ are the theoretical (first $7$ columns) 
and experimental (last column) particle ratios \cite{star-2009}. The temperatures and
baryonic chemical potentials obtained from the statistical model in \cite{star-2009} are
also given in the last columns of all tables. 

In figs. 1-$a/b$, 2-$a/b$, 3-$a/b$ and 4-$a/b$  we plot the
experimental and theoretical ratios for $B=0$ and the best $B\neq0$.
In figs. 1-$c$, 2-$c$, 3-$c$ and 4-$c$ we show the $\chi^2$ behavior for $B=0$ and for the best $B\neq0$.
In figs. 1-$d$, 2-$d$, 3-$d$ and 4-$d$ we show the $\chi^2$ behavior
for the different magnetic fields.
One can notice that the best fitting is generally obtained for magnetic fields around 6 $m_\pi^2$,
a little higher than what is expected for RHIC collisions (5 $m_\pi^2$).

Our results show that, even for the free Fermi and Boson gas models, a strong magnetic field plays an important role.
The inclusion of the magnetic field improves the data fit up to a field of the order of $B=10^{19}$ G.
For stronger magnetic fields, it becomes worse again.
This behavior is easily observed in tables 1 to 4 and in figs.1-d to 4-d.
It is worth pointing out how the "fireball" radius $R$ and the total
density $\rho$ vary with the magnetic field in a systematic way:
$R$ and $\rho$ practically do not change between $B=0$ and $B=10^{18}$ G,
but when the field increases even further, the density increases and the radius decreases.
This behavior is common to all collision cases studied.
This huge jump in the density explains why the ratios get worse for a
magnetic field of the order of $B=10^{20}$ G, for which
the densities are much higher than what is expected in a heavy ion collision. 

Our model gives a good description for the particle/antiparticle
ratios, but fails to describe the relation between baryons and mesons.
This occurs because our model produces too many mesons (especially
pions) as shown explicitly in the particle densities. In all collision
types our model presents 
a baryon density ($\rho_B$) with more than 30\% of $\Delta$ baryons and
a meson density ($\rho_M$) with more than 60\% of $\pi$ ($\rho_\pi$).
The relative percent deviations in the particle yields show
clearly that some results improve considerably when the magnetic
field is considered, while others remain unaltered or even get
slightly worse. However, our figures also show that the behavior 
of the $\chi^2$ changes drastically with the addition of the 
magnetic field  and that the temperature and chemical potentials 
calculated with the statistical model lie within
the  $3-\sigma$ confidence ellipse obtained for the best $\chi^2$ in
some cases, but
they are always outside the confidence ellipses obtained with zero
magnetic field.

 We would like to comment that when we first started these
  calculations, we were not aware of references \cite{skokov,deng} and
we used data obtained for low centralities, i.e., low impact
parameters. In that case, the minimum $\chi^2$ was generally smaller
than the ones shown in this work and we believe this was so because of
the larger error bars accompanying data at low centralities.

Further improvements on the presented calculations are under
investigation, namely, the inclusion of electric fields at low
impact parameters and the variation of both electric and magnetic
fields with the number of participants in the collisions. Moreover, 
we are working on the inclusion of the anomalous magnetic moments
and in the description of pion-pion interactions.
We next intend to repeat these calculations for the ALICE/LHC data for
the  future Au$+$Au runs with all these improvements, so that our
results become more realistic.

\acknowledgments

This work was partially supported by CNPq, CAPES and FAPESC (Brazil). 
We thank very fruitful discussions with Dr. Celso Camargo de Barros and
Dr. Sidney dos Santos Avancini.

\begin{landscape}
\begin{table}
\begin{tabular}{|l|ccccccc|c|}
\hline
$B$ ($\times10^{19}$G) &0&0.1&0.5&1&\bf2& 5&10&\multirow{2}{*}{STAR/RHIC\cite{star-2009}} \\
$eB$ ($m_\pi^2$)       &0&0.3&1.5&3&\bf6&15&30& \\
\hline
$\pi^-/\pi^+$ $(\Delta_\%)$&1.000 (0.30\%)&1.000&1.000&1.000&\bf1.000 (0.30\%)&1.000&1.000&1.003$\pm$0.044\\
$K^-/K^+$                  &0.993 (1.23\%)&0.993&0.986&0.980&\bf0.971 (0.99\%)&0.966&0.939&0.981$\pm$0.049\\
$\bar{p}/p$                &0.842 (0.13\%)&0.843&0.834&0.839&\bf0.837 (0.69\%)&0.836&0.776&0.843$\pm$0.048\\
$K^-/\pi^-$                &0.169 (32.7\%)&0.169&0.175&0.177&\bf0.178 (40.1\%)&0.182&0.218&0.127$\pm$0.010\\
$\bar{p}/\pi^-$            &0.019 (77.8\%)&0.019&0.026&0.033&\bf0.037 (55.7\%)&0.036&0.046&0.084$\pm$0.007\\
$K^+/\pi^+$                &0.170 (30.5\%)&0.170&0.177&0.181&\bf0.183 (40.9\%)&0.189&0.232&0.130$\pm$0.011\\
$p/\pi^+$                  &0.022 (77.9\%)&0.023&0.031&0.039&\bf0.044 (55.6\%)&0.043&0.059&0.100$\pm$0.013\\
\hline
$T$ (MeV)                               &    124&    125&   138&   152&  \bf170&  194&  199& 157.9$\pm$3.9\\
$\mu_B$ (MeV)                           &     11&     11&    13&    14&   \bf16&   19&   20& 14.1$\pm$4.2\\
$\mu_S$ (MeV)                           &  0.758&  0.787&  1.43&  2.15& \bf3.28& 4.82& 5.08& \\
$\mu_{I3}$ (MeV)                        & -0.653& -0.660&-0.949& -1.22&\bf-1.76&-3.24&-5.00& \\
$\chi^2/ndf$                            &   42.5&   52.5&  46.0&  39.8& \bf35.6& 39.7& 47.9& \\
$R$ (fm)                                &   66.2&   64.6&  45.5&  34.0& \bf24.5& 16.8& 14.8& \\
$\rho_B$ ($\times10^{-2}$fm$^{-3}$)     & 0.0325& 0.0349& 0.100& 0.239&\bf0.643& 2.00& 2.91& \\
$\rho_\Delta$ ($\times10^{-2}$fm$^{-3}$)&0.00813&0.00878&0.0276&0.0717&\bf0.211&0.732& 1.14& \\
$\rho_M$ (fm$^{-3}$)                    & 0.0918& 0.0960& 0.167& 0.293&\bf0.582& 1.51& 2.73& \\
$\rho_\pi$ (fm$^{-3}$)                  & 0.0636& 0.0664& 0.109& 0.183&\bf0.355&0.942& 1.83& \\
$\epsilon$ (MeV/fm$^{3}$)               &   55.4&   58.3&   111&   212&  \bf456& 1203& 1820& \\
$P$ (MeV/fm$^{3}$)                      &   11.2&   11.8&  22.6&  43.7& \bf97.4&  283&  490& \\
$s$ (MeV/fm$^{3}$)                      &  0.537&  0.561& 0.967&  1.68& \bf3.26& 7.66& 11.6& \\
\hline
\end{tabular}
\caption{Au$+$Au (70-80\%) $\sqrt{s_{NN}}$ = 200 GeV.
Results obtained for different values of the magnetic field.} 
\end{table}
\end{landscape}

\begin{figure}
\includegraphics[width=.5\textwidth]{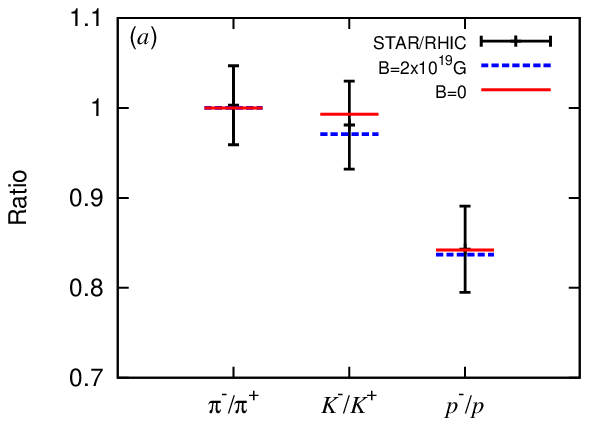}\includegraphics[width=.5\textwidth]{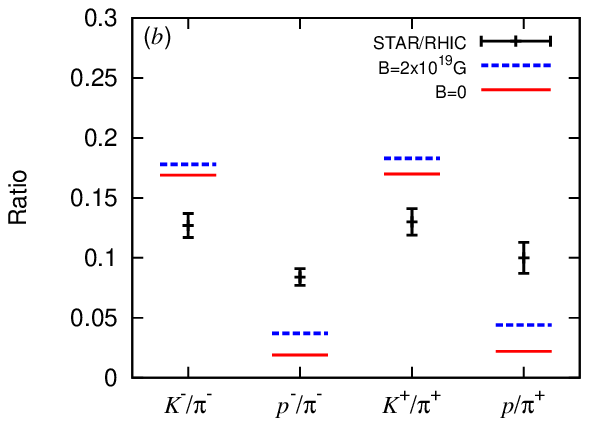}\\
\includegraphics[width=.5\textwidth]{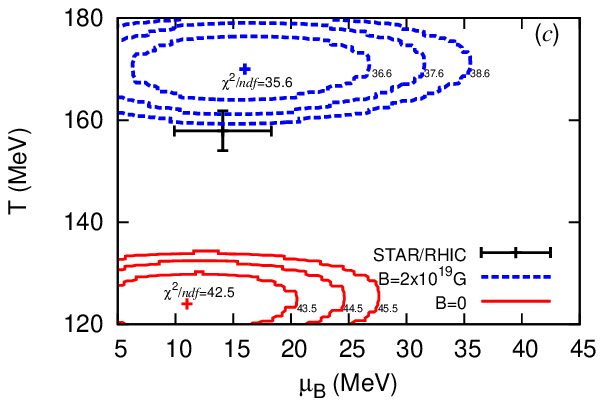}\includegraphics[width=.5\textwidth]{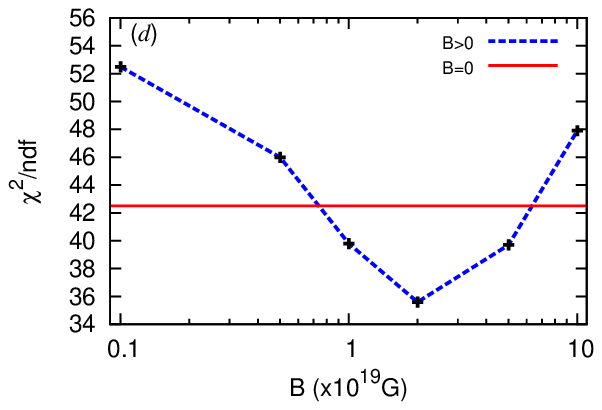}\\
\caption{Au$+$Au (70-80\%) collision at $\sqrt{s_{NN}}$ = 200 GeV.
$(a)$ particle/antiparticle ratios. $(b)$ mixed ratios. $(c)$ and $(d)$ $\chi^2$ behavior.}
\end{figure}

\begin{landscape}
\begin{table}
\begin{tabular}{|l|ccccccc|c|}
\hline
$B$ ($\times10^{19}$G) &0&0.1&0.5&1&\bf2& 5&10&\multirow{2}{*}{STAR/RHIC\cite{star-2009}} \\
$eB$ ($m_\pi^2$)       &0&0.3&1.5&3&\bf6&15&30& \\
\hline
$\pi^-/\pi^+$ $(\Delta_\%)$&1.000 (0.40\%)&1.000&1.000&1.000&\bf1.000 (0.40\%)&1.000&1.000&0.996$\pm$0.066\\
$K^-/K^+$                  &0.985 (2.32\%)&0.984&0.977&0.967&\bf0.956 (0.71\%)&0.947&0.952&0.963$\pm$0.050\\
$\bar{p}/p$                &0.793 (0.35\%)&0.783&0.788&0.787&\bf0.792 (0.25\%)&0.794&0.792&0.790$\pm$0.043\\
$K^-/\pi^-$                &0.196 (40.3\%)&0.196&0.195&0.195&\bf0.194 (38.53\%)&0.199&0.201&0.140$\pm$0.018\\
$\bar{p}/\pi^-$            &0.027 (67.6\%)&0.027&0.033&0.040&\bf0.045 (44.9\%)&0.045&0.036&0.082$\pm$0.010\\
$K^+/\pi^+$                &0.199 (38.2\%)&0.200&0.200&0.202&\bf0.203 (40.9\%)&0.210&0.211&0.144$\pm$0.016\\
$p/\pi^+$                  &0.034 (67.4\%)&0.035&0.042&0.050&\bf0.057 (44.6\%)&0.057&0.046&0.103$\pm$0.012\\
\hline
$T$ (MeV)                               &   133&   134&   145&  159&  \bf178&  206&  217& 159$^{+11}_{-7}$\\
$\mu_B$ (MeV)                           &    16&    17&    18&   20&   \bf22&   26&   29& 19.9$\pm$4.9\\
$\mu_S$ (MeV)                           &  1.52&  1.67&  2.39& 3.55& \bf5.07& 7.38& 8.42& \\
$\mu_{I3}$ (MeV)                        & -1.10& -1.18& -1.41&-1.79&\bf-2.41&-4.26&-6.92& \\
$\chi^2/ndf$                            &  17.2&  21.2&  17.9& 14.8& \bf12.7& 14.2& 18.1& \\
$R$ (fm)                                &  47.1&  45.2&  35.5& 26.9& \bf19.8& 13.4& 11.1& \\
$\rho_B$ ($\times10^{-2}$fm$^{-3}$)     &0.0899& 0.102& 0.211&0.486& \bf1.22& 3.95& 6.92& \\
$\rho_\Delta$ ($\times10^{-2}$fm$^{-3}$)&0.0243&0.0273&0.0610&0.151&\bf0.415& 1.50& 2.82& \\ 
$\rho_M$ (fm$^{-3}$)                    & 0.127& 0.132& 0.207&0.352& \bf0.693& 1.84& 3.49& \\
$\rho_\pi$ (fm$^{-3}$)                  &0.0813&0.0845& 0.127&0.208&\bf0.398& 1.07& 2.16& \\
$\epsilon$ (MeV/fm$^{3}$)               &  86.4&  90.6&   152&  282&  \bf608& 1704& 2863& \\
$P$ (MeV/fm$^{3}$)                      &  17.0&  17.8&  30.2& 56.6&  \bf126&  385&  729& \\
$s$ (MeV/fm$^{3}$)                      & 0.777& 0.809&  1.26& 2.13& \bf4.12& 10.1& 16.5& \\
\hline
\end{tabular}
\caption{Au$+$Au (58-85\%) collision at $\sqrt{s_{NN}}$ = 130 GeV.
Results obtained for different values of the  magnetic field.}
\end{table}
\end{landscape}

\begin{figure}
\includegraphics[width=.5\textwidth]{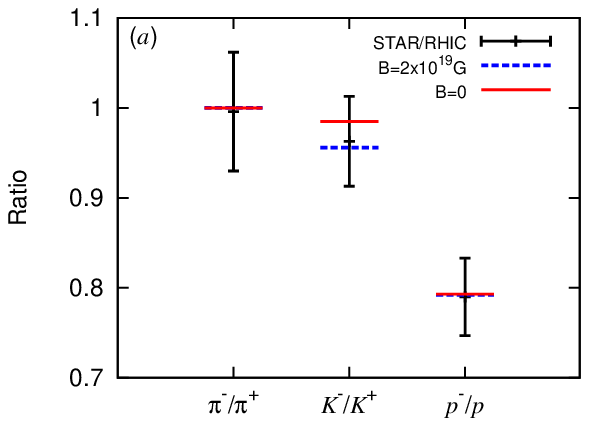}\includegraphics[width=.5\textwidth]{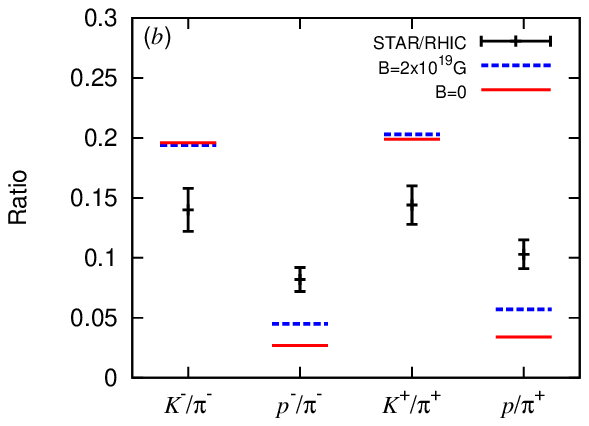}\\
\includegraphics[width=.5\textwidth]{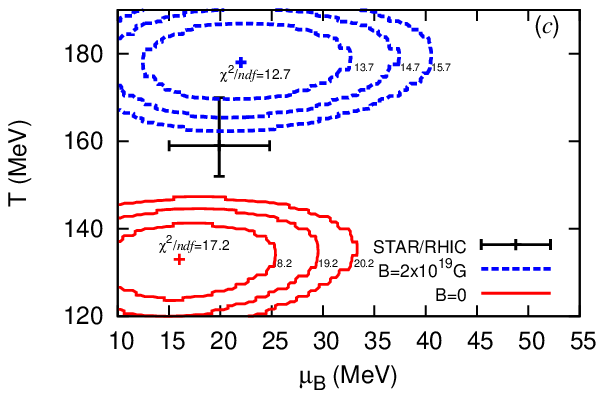}\includegraphics[width=.5\textwidth]{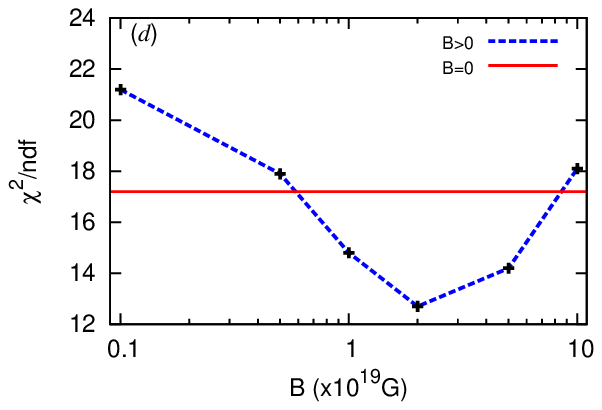}\\
\caption{Au$+$Au (58-85\%) collision at $\sqrt{s_{NN}}$ = 130 GeV.
$(a)$ particle/antiparticle ratios. $(b)$ mixed ratios. $(c)$ and $(d)$ $\chi^2$ behavior.}
\end{figure}

\begin{landscape}
\begin{table}
\begin{tabular}{|l|ccccccc|c|}
\hline
$B$ ($\times10^{19}$G) &0&0.1&0.5&1&\bf2& 5&10&\multirow{2}{*}{STAR/RHIC\cite{star-2009}} \\
$eB$ ($m_\pi^2$)       &0&0.3&1.5&3&\bf6&15&30& \\
\hline
$\pi^-/\pi^+$ $(\Delta_\%)$&1.000 (1.19\%)&1.000&1.000&1.000&\bf1.000 (1.19\%)&1.000&1.000&1.012$\pm$0.031\\
$K^-/K^+$                  &0.992 (5.97\%)&0.991&0.982&0.971&\bf0.958 (2.34\%)&0.959&0.972&0.936$\pm$0.036\\
$\bar{p}/p$                &0.621 (0.27\%)&0.624&0.623&0.625&\bf0.625 (0.26\%)&0.623&0.618&0.623$\pm$0.047\\
$K^-/\pi^-$                &0.128 (17.0\%)&0.128&0.130&0.132&\bf0.132 (20.9\%)&0.131&0.129&0.109$\pm$0.009\\
$\bar{p}/\pi^-$            &0.008 (87.1\%)&0.008&0.011&0.014&\bf0.016 (74.7\%)&0.013&0.010&0.063$\pm$0.007\\
$K^+/\pi^+$                &0.129 (8.99\%)&0.129&0.133&0.136&\bf0.138 (16.6\%)&0.137&0.132&0.118$\pm$0.010\\
$p/\pi^+$                  &0.013 (87.1\%)&0.013&0.018&0.023&\bf0.026 (74.7\%)&0.021&0.016&0.101$\pm$0.013\\
\hline
$T$ (MeV)                               &    111&    112&   123&    135&  \bf149&  162&  193& 154$^{+8}_{-6}$\\
$\mu_B$ (MeV)                           &     27&     27&    30&     33&   \bf37&   42&   45& 37.7$\pm$6.5\\
$\mu_S$ (MeV)                           &   1.04&   1.09&  1.97&   3.24& \bf5.06& 6.93& 7.85& \\
$\mu_{I3}$ (MeV)                        &  -1.17&  -1.20& -1.75&  -2.49&\bf-3.83&-7.28&-11.3& \\
$\chi^2/ndf$                            &   23.0&   28.5&  26.0&   23.7& \bf22.4& 24.9& 27.5& \\
$R$ (fm)                                &   70.4&   68.4&  48.6&   35.4& \bf25.7& 19.3& 17.3& \\
$\rho_B$ ($\times10^{-2}$fm$^{-3}$)     & 0.0269& 0.0295&0.0821&  0.212&\bf0.552& 1.31& 1.81& \\
$\rho_\Delta$ ($\times10^{-2}$fm$^{-3}$)&0.00584&0.00643&0.0199& 0.0566&\bf0.162&0.430&0.648& \\
$\rho_M$ (fm$^{-3}$)                    & 0.0557& 0.0587& 0.103&  0.183&\bf0.361&0.871& 1.59& \\
$\rho_\pi$ (fm$^{-3}$)                  & 0.0429& 0.0451&0.0758&  0.131&\bf0.257&0.646& 1.24& \\
$\epsilon$ (MeV/fm$^{3}$)               &   28.1&   29.8&  54.4&    102&  \bf206&  444&  687& \\
$P$ (MeV/fm$^{3}$)                      &   5.97&   6.33&  11.9&   22.8& \bf48.9&  122&  214& \\
$s$ (MeV/fm$^{3}$)                      &  0.307&  0.322& 0.539&  0.922& \bf1.71& 3.49& 5.52& \\
\hline
\end{tabular}
\caption{Au$+$Au (70-80\%) collision at $\sqrt{s_{NN}}$ = 62.4 GeV.
Results obtained for different values of the magnetic field.}
\end{table}
\end{landscape}

\begin{figure}
\includegraphics[width=.5\textwidth]{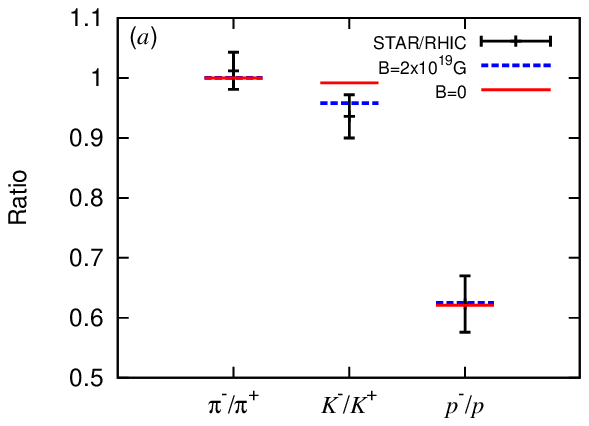}\includegraphics[width=.5\textwidth]{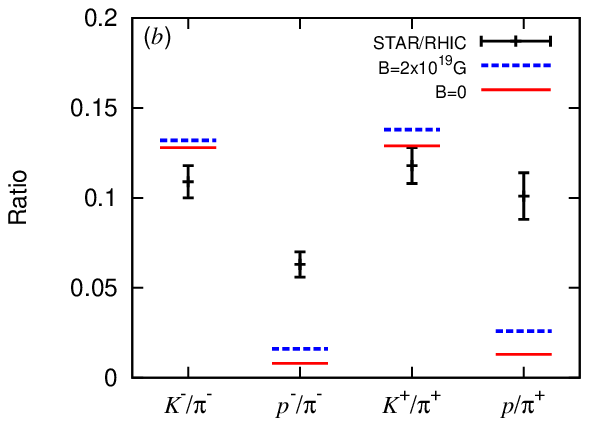}\\
\includegraphics[width=.5\textwidth]{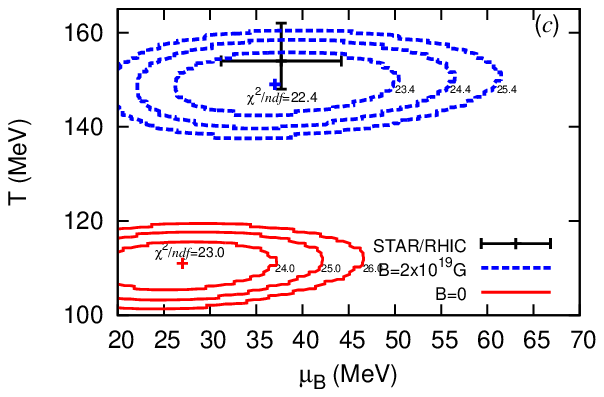}\includegraphics[width=.5\textwidth]{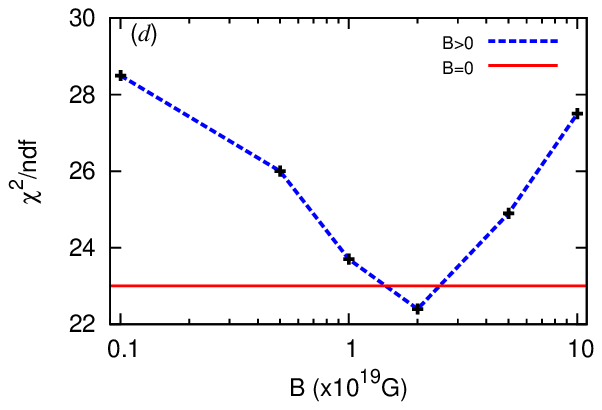}\\
\caption{Au$+$Au (70-80\%) collision at $\sqrt{s_{NN}}$ = 62.4 GeV.
$(a)$ particle/antiparticle ratios. $(b)$ mixed ratios. $(c)$ and $(d)$ $\chi^2$ behavior.}
\end{figure}

\begin{landscape}
\begin{table}
\begin{tabular}{|l|ccccccc|c|}
\hline
$B$ ($\times10^{19}$G) &0&0.1&0.5&1&\bf2& 5&10&\multirow{2}{*}{STAR/RHIC\cite{star-2009}} \\
$eB$ ($m_\pi^2$)       &0&0.3&1.5&3&\bf6&15&30& \\
\hline
$\pi^-/\pi^+$ $(\Delta_\%)$&1.000 (0.79\%)&1.000&1.000&1.000&\bf1.000 (0.79\%)&1.000&1.000&1.008$\pm$0.042\\
$K^-/K^+$                  &0.996 (1.98\%)&0.996&0.992&0.988&\bf0.981 (0.43\%)&0.981&0.987&0.977$\pm$0.037\\
$\bar{p}/p$                &0.843 (0.18\%)&0.844&0.844&0.847&\bf0.843 (0.22\%)&0.842&0.841&0.841$\pm$0.067\\
$K^-/\pi^-$                &0.137 (14.3\%)&0.137&0.140&0.143&\bf0.146 (21.8\%)&0.146&0.142&0.120$\pm$0.011\\
$\bar{p}/\pi^-$            &0.011 (86.3\%)&0.012&0.015&0.020&\bf0.023 (71.8\%)&0.019&0.014&0.082$\pm$0.010\\
$K^+/\pi^+$                &0.138 (11.0\%)&0.138&0.141&0.145&\bf0.149 (20.1\%)&0.149&0.144&0.124$\pm$0.012\\
$p/\pi^+$                  &0.013 (86.4\%)&0.014&0.018&0.024&\bf0.027 (71.0\%)&0.023&0.017&0.098$\pm$0.014\\
\hline
$T$ (MeV)                               &    114&    115&    126&   139&   \bf155&  170&  171& 159$^{+10}_{-7}$\\
$\mu_B$ (MeV)                           &     10&     10&     11&    12&    \bf14&   16&   17& 16.5$\pm$6.5\\
$\mu_S$ (MeV)                           &  0.443&  0.464&  0.807&  1.32&  \bf2.18& 3.00& 3.26&\\
$\mu_{I3}$ (MeV)                        & -0.467& -0.475& -0.671&-0.939& \bf-1.49&-2.80&-4.33& \\
$\chi^2/ndf$                            &   18.2&   22.5&   20.5&  18.6&  \bf17.5& 19.4& 21.7& \\
$R$ (fm)                                &   71.5&   69.5&   50.2&  36.4&  \bf25.6& 18.9& 17.1& \\
$\rho_B$ ($\times10^{-2}$fm$^{-3}$)     & 0.0130& 0.0141& 0.0375&0.0982& \bf0.283&0.700&0.944& \\
$\rho_\Delta$ ($\times10^{-2}$fm$^{-3}$)&0.00292&0.00320&0.00936&0.0270&\bf0.0859&0.237&0.344& \\
$\rho_M$ (fm$^{-3}$)                    & 0.0627& 0.0660&  0.113& 0.204& \bf0.415& 1.00& 1.80& \\
$\rho_\pi$ (fm$^{-3}$)                  & 0.0472& 0.0495& 0.0818& 0.142& \bf0.283&0.715& 1.36& \\
$\epsilon$ (MeV/fm$^{3}$)               &   33.0&   34.9&   62.9&   121&   \bf259&  573&  857& \\
$P$ (MeV/fm$^{3}$)                      &   6.93&   7.33&   13.5&  26.6&  \bf59.6&  151&  258& \\
$s$ (MeV/fm$^{3}$)                      &  0.350&  0.367&  0.606&  1.06&  \bf2.06& 4.25& 6.52& \\
\hline
\end{tabular}
\caption{$d+$Au (40-100\%) collision at $\sqrt{s_{NN}}$ = 200 GeV.
Results obtained for different values of the magnetic field.}
\end{table}
\end{landscape}

\begin{figure}
\includegraphics[width=.5\textwidth]{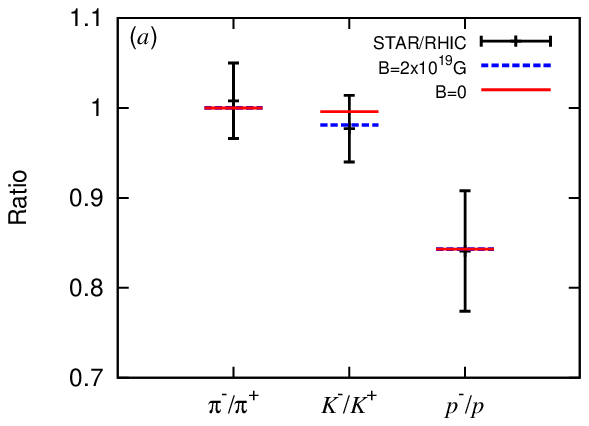}\includegraphics[width=.5\textwidth]{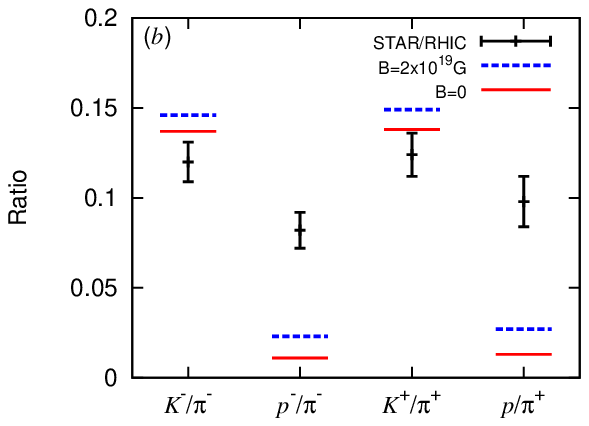}\\
\includegraphics[width=.5\textwidth]{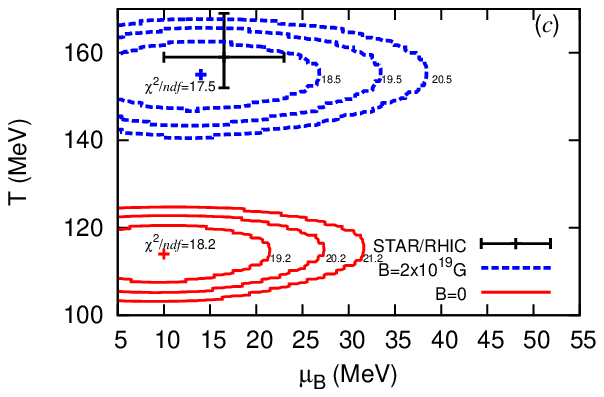}\includegraphics[width=.5\textwidth]{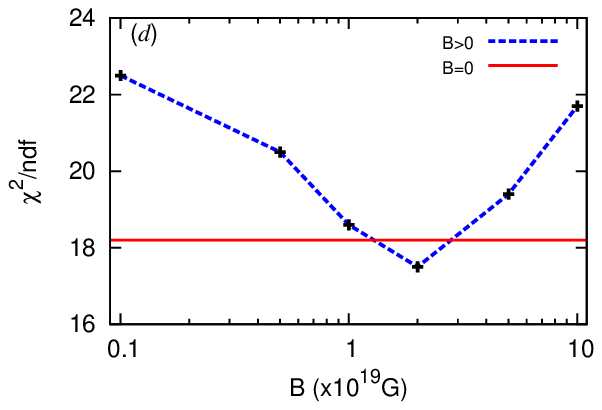}\\
\caption{$d+$Au (40-100\%) collision at $\sqrt{s_{NN}}$ = 200 GeV. $(a)$ particle/antiparticle ratios.
$(b)$ mixed ratios. $(c)$ and $(d)$ $\chi^2$ behavior.}
\end{figure}

\end{document}